\documentclass[twocolumn,showpacs,preprintnumbers]{revtex4}

\usepackage{amsmath,amssymb}

\usepackage{graphicx}

\usepackage{bm}
\usepackage[latin1]{inputenc}

\newcommand{\bra}[1]{\langle\left.{#1}\right|}
\newcommand{\ket}[1]{\left|{#1}\right.\rangle}
\newcommand{\xpct}[1]{\langle{#1}\rangle}    

\begin{document}

\title{Edge-localized states in quantum one-dimensional lattices}

\author{Ricardo A. Pinto} 
\author{Masudul Haque} 
\author{Sergej Flach}
\affiliation{Max-Planck-Institut f\"ur Physik komplexer Systeme, N\"othnitzer
Str. 38, 01187 Dresden, Germany}

\date{\today}

\begin{abstract}

In one-dimensional quantum lattice models with open boundaries, we find and
study localization at the lattice edge.  We show that edge-localized
eigenstates can be found in both bosonic and fermionic systems, specifically,
in the Bose-Hubbard model with on-site interactions and in the spinless
fermion model with nearest-neighbor interactions.  We characterize the
localization through spectral studies via numerical diagonalization and
perturbation theory, through considerations of the eigenfunctions, and through
the study of explicit time evolution.  We concentrate on few-particle systems,
showing how more complicated edge states appear as the number of particles is
increased.

\end{abstract}

\pacs{63.20.Pw, 63.20.Ry, 63.22.+m, 03.65.Ge}

\maketitle

\section{Introduction}

Experimental techniques in the fields of mesoscopics and ultracold atoms have
advanced to the point where it is feasible to explore the physics of few-boson
and few-fermion systems, in particular in lattice systems which have
traditionally been the basis for many-particle physics studies.
The study of a few quantum particles in lattice situations provides surprises
and unexpected phenomena quite distinct from issues in bulk condensed matter
physics, whose focus is on \emph{many}-particle ground states, and from atomic
or optical physics, where lattice systems are not very common.
For example, a recent experiment has explored repulsive binding of boson pairs
in a one-dimensional optical lattice
\cite{GrimmZoller_pair-binding_Nature2006}.  
This is therefore the appropriate time to investigate further intricate and
non-intuitive phenomena involving a few quantum particles in lattice systems.
In this article, we consider interacting bosons or fermions in one-dimensional
finite lattices, and present studies of \emph{localization} at the lattice
edge.  
We characterize the edge-localized states, and distinguish them from bound
non-localized states, through analyses of the energy spectrum and band
structure, density profiles of eigenstates, and dynamics.

For bosons, we use the well-known Bose-Hubbard model
\cite{FisherWeichnmanFisher_PRB89}, which has attracted a great deal of
attention in the last decade due to its relevance to describing laser-cooled
bosonic atoms subjected to an optical-lattice potential
\cite{JakschCiracGardinerZoller_PRL98, GreinerBloch_Nature2002}.
Localization in the Bose-Hubbard model is of particular interest because its
large-boson limit can for many purposes be approximated by the discrete
nonlinear Schr\"odinger (DNLS) equation, which displays a host of localization
phenomena.  Nonlinearity allows time-periodic and spatially-localized
solutions of the DNLS (and other lattice nonlinear differential equations)
known as \emph{intrinsic localized modes} or \emph{discrete breathers}
\cite{FlachWillis_PhysRep1998, physicstoday, Aubry_PhysicaD1997, Sievers,
Aubry_review_PhysicaD2006}.
More relevant to the present work, the DNLS on finite lattices also possesses
edge-localized modes, sometimes called discrete surface solitons
\cite{MakrisSuntsovChristodoulidesStegemanHache_OptLett2005,
MolinaVicencioKivshar_OptLett2006, Suntsov2006PRL96_and_others}.  It is therefore expected that the
large-boson limit of the open-boundary Bose-Hubbard model will possess
eigenstates in which the bosons are localized at the edge.  In this Article,
we pose the edge-localization question in the extreme opposite limit of a
\emph{few} quantum particles, where the mean-field approximation (DNLS
equation) cannot \emph{a priori} be expected to provide the correct intuition.
The answer turns out to be subtle --- this phenomenon is not present for the
case of two particles, but appears when the particle number is three or more,
as follows from numerical studies in Ref.~\cite{Pouthier_PRB2007}.

Remarkably, we find that the phenomenon is not restricted to bosons, but also
happens in other quantum lattice models.  For example, the spinless fermion
model with nearest-neighbor interactions, sometimes known as the $t$-$V$
model, has similar behavior, \emph{i.e.}, it possesses no edge states with two
fermions, but does have such localized eigenstates with three fermions.  Since
this model does not allow multiple occupancies, localization in this case
refers to a sequence of neighboring sites being occupied, rather than all the
particles clustered in one site as in the case of bosons.  Although the
current work will focus on the two models mentioned above, our finding
indicates that edge-localization may well be a generic phenomenon in quantum
finite lattice systems.

Localization in quantum models generally does not appear in simple
Hamiltonians, but instead requires disorder \cite{Anderson_localizn_PR1958} or
impurities breaking the translation invariance.  (This is in contrast to
lattice differential equations, \emph{e.g.}, the DNLS equation, where the
interplay of nonlinearity and spatial discreteness is sufficient to create
localization.)  It is therefore of significant theoretical interest to explore
the simple mechanism for quantum localization that is studied here, requiring
only open boundary conditions in an interacting lattice Hamiltonian.  We also
note that localization due to impurities/disorder is generally a
single-particle effect, while the mechanism we present is a \emph{collective}
phenomenon since it requires at least three particles.  
%

While we focus on small numbers of bosons or fermions, the basic message is
that edge states exist for any number of particles larger than two.  We give
an explanation based on perturbation theory.  We also provide some hints
toward the many-particle situation, by showing that the four-boson case has an
additional type of edge-localized eigenstate in addition to the obvious
generalization.

\emph{Model Hamiltonians}.
We consider one-dimensional lattices with $L$ sites subject to open boundary
conditions.  The Bose-Hubbard model Hamiltonian is
\begin{equation}\label{eq:Bose-Hubbardhamiltonian}
\hat{H}_{\rm BH} ~=~ - \frac{\gamma}{2}\sum_{j=1}^L
\hat{a}_j^{\dagger}\hat{a}_j^{\dagger}\hat{a}_j\hat{a}_j ~-~ t \sum_{j=1}^{L-1}
\left( \hat{a}_j^{\dagger}\hat{a}_{j+1}  + \hat{a}_{j+1}^{\dagger}\hat{a}_j \right).
\end{equation}
Here $\hat{a}$ and $\hat{a}^{\dagger}$ are second-quantized bosonic operators.  The
first and second term describe respectively on-site attractive ($\gamma>0$) interactions
and nearest-neighbor hopping.  
%
%
%
The edge-localization physics is almost unchanged in the case of repulsive
interactions.


The model for spinless fermions is described by
\begin{equation}  \label{eq:tVmodelHamiltonian}
\hat{H}_{\rm sf} ~=~
V\sum_{j=1}^{L-1}\hat{c}_j^{\dagger}\hat{c}_{j+1}^{\dagger}\hat{c}_{j+1}\hat{c}_j
~-~ t\sum_{j=1}^{L-1} \left( \hat{c}_j^{\dagger}\hat{c}_{j+1}
 ~+~ \hat{c}_{j+1}^{\dagger}\hat{c} \right) ,
\end{equation}
where $V$ is the (repulsive) nearest-neighbor interaction strength, and
$\hat{c}_j$ and $\hat{c}_j^{\dagger}$ are second-quantized fermionic
operators.



\section{Bose-Hubbard chain with two and three bosons}\label{twobosonstates}

\begin{figure}
\includegraphics[width=0.98\columnwidth]{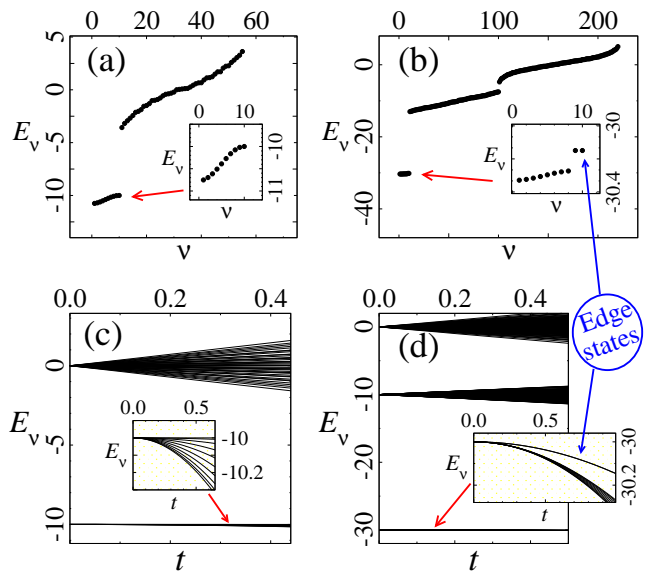}
\caption{\label{fig_BH-spectra} 
(Color online.)  Energy spectrum of 10-site Bose-Hubbard chain with
$\gamma=10$.  Left panels (a,c) show the case of two bosons ($n=2$), right
panels (b,d) show $n=3$.  Top panels (a,b) plot ordered energies against index
$\nu$, for fixed hopping parameter $t=1$. Lower panels (c,d) plot energies
against hopping strength $t$.  Insets focus on the lowest (breather) band,
showing that two edge-localized states separate out for three bosons (b,d) but
not for two bosons (a,c).
}
\end{figure} 

We will first describe the cases of two bosons and of three bosons in the
Bose-Hubbard chain, and show how the two situations differ by virtue of the
latter having edge states.  
Some of the physics described in this section appears in
Ref.~\cite{Pouthier_PRB2007}.

\emph{Strong interactions}. ---
We first focus on large values of $\gamma$, which is the most relevant
parameter regime for localization physics. 

The Hilbert space size (number of basis states) for $n$ bosons in $L>n$ sites
is $d_n= \begin{pmatrix}L+n-1\\n\end{pmatrix}$;
in particular $d_2= L(L+1)/2$ and $d_3=L(L+1)(L+2)/6$.  Figure \ref{fig_BH-spectra}
displays spectral properties and band structure, \emph{i.e.}, the distribution
of the $d_n$ eigenenergies.  We label the eigenstates from low to high
energies with the label $\nu$ running from 1 to $d_n$.

The most prominent feature of the large-$\gamma$ spectrum is the band
structure.  For the $n$-boson system there are $p(n)$ bands, where $p(n)$ is
the number of integer partitions of $n$.
In the two-boson case (Figure \ref{fig_BH-spectra}-a and \ref{fig_BH-spectra}-c), there are two bands. 
In the upper band around zero energy, the dominant contributions come from
configurations where the bosons do not share the same site.
The lower band contains the $L$ two-boson bound states, dominated by linear
combinations of configurations where the bosons sit on the same site.  These
are the quantum analogs of classical discrete breather solutions.  This band
is thus called the quantum breather band or the soliton band
\cite{EilbeckPhysicaD78}.  In these states, the separation probability of the
bosons decays exponentially with distance
\cite{Eilbeck03,Scott1,NguenangPRB75}.

\begin{figure}
\includegraphics[width=0.9\columnwidth]{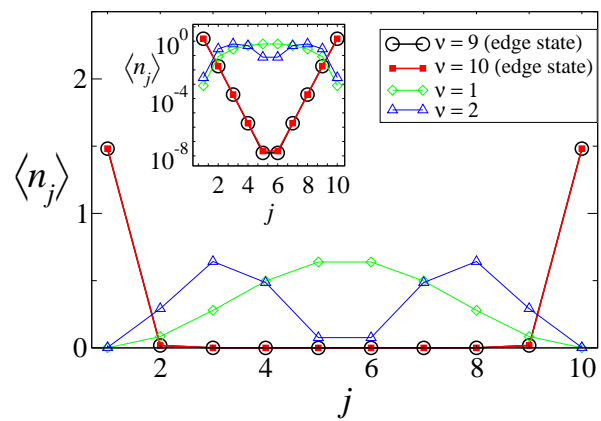}
\caption{\label{3bosonedgestates}
(Color online.)  Spatial profile of site occupancies for several eigenstates
of the three-boson Bose-Hubbard chain.  Inset shows the same plot in semilog
scale, the linear behavior indicating exponential localization in the edge
states. Here $L=10$ and $\gamma=10$.}
\end{figure} 

In the three-boson case (Figures \ref{fig_BH-spectra}-b and \ref{fig_BH-spectra}-d), the
spectrum contains three energy bands. The lowest-energy band is formed by the
$L$ three-boson bound states (three-boson breather band), where there is a
high probability of finding the three bosons at the same lattice site. The
second band from below is formed by $L(L-1)$ ``2+1-boson states", where there
is a high probability of finding two bosons at the same lattice site with the
third boson elsewhere. Finally, the third band is the three-boson continuum,
whose $L(L-1)(L-2)/6$ eigenstates are characterized by having the three bosons
in different sites.

The edge states can be identified by zooming onto the three-boson bound state band
(insets of Figures \ref{fig_BH-spectra}-b and \ref{fig_BH-spectra}-d).  We note that two
states stand out from the rest of the band, with larger splitting.  These are
the edge states, as we demonstrate further below.  Because of reflection
symmetry, the dominant contributors to the two eigenstates are not the
left-edge and right-edge states ($\ket{E_{l}} = \ket{3000...}$ and
$\ket{E_{r}} = \ket{...0003}$) directly.  Instead, the eigenstates are
dominated by the linear combinations $\tfrac{1}{\sqrt{2}}\Bigl(\ket{E_{l}}\pm\ket{E_{r}}\Bigr)$.
The remaining $(L-2)$ eigenstates of the breather band are dominated by linear
combinations of the remaining $(L-2)$ three-boson bound state configurations,
$\ket{03000...}$, $\ket{00300...}$, ... $\ket{...0030}$.  These $(L-2)$ states
do not mix with the two edge states because of the energy splitting.

Note that the energy-separated edge-localized states are not present in the
two-boson case (insets of Figures \ref{fig_BH-spectra}-a and \ref{fig_BH-spectra}-c).  

To see the localized nature of the edge states, in Figure
\ref{3bosonedgestates} we plot density profiles (site occupancies),
\emph{i.e.}, expectation values of boson number at each lattice site,
$\xpct{n_j}_{\nu} = \langle
\chi_{\nu}|\hat{a}_j^{\dagger}\hat{a}_j|\chi_{\nu}\rangle$,
$|\chi_{\nu}\rangle$ being an eigenstate.
Figure \ref{3bosonedgestates} shows $\langle{n_j}\rangle$ against $j$ for the
edge states, and also for two other eigenstates (non-edge breather states) for
comparison. We see that the edge states are exponentially localized at the
edges of the lattice, as seen through the linear behavior of
$\ln\langle{n_j}\rangle$ in the inset.

\begin{figure}
\includegraphics[width=0.8\columnwidth]{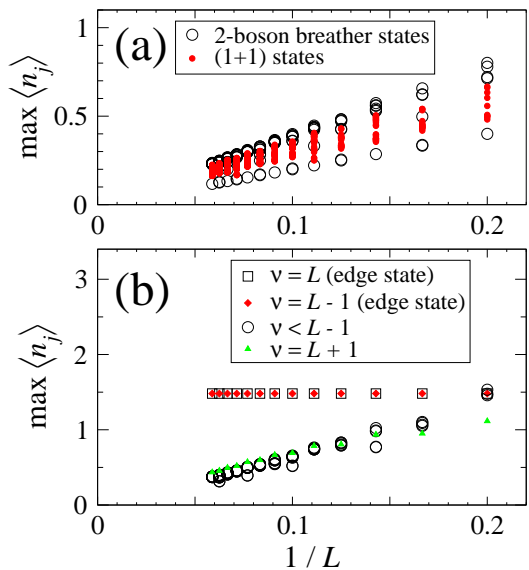}
\caption{\label{fig_size-dependence} 
(Color online.)  Size-dependence of $\max\xpct{n_j}$ for (a) all eigenstates
of the two-boson Bose-Hubbard chain, and (b) several eigenstates of the
three-boson Bose-Hubbard chain labeled by $\nu$. Here $\gamma=10$, and $L$
goes from $L=5$ to $L=17$. }
\end{figure} 

The localization is further demonstrated through the scaling of the largest
site occupancies, $\max\langle n_j\rangle$, with system size $L$ (Figure
\ref{fig_size-dependence}).  For eigenstates that are localized, this quantity
will be independent of the system size, whereas for extended states it should
be a linear function of $1/L$ because the bosons are spread across $L$ sites
in extended states.  In the two-bosons case (Figure
\ref{fig_size-dependence}-a) all eigenstates are extended, thus
$\max\xpct{n_j}$ depends linearly on $1/L$ for all of them.  In Figure
\ref{fig_size-dependence}-b for three bosons, most states also have
$\max\xpct{n_j}$ varying as ${\sim}L^{-1}$, except for two states for which
$\max\xpct{n_j}$ are independent of $L$.  This flat set of points in Figure
\ref{fig_size-dependence}-b clearly illustrates the localization phenomenon.

We have described the spectrum for the attractive Bose-Hubbard model.  The
spectrum for the repulsive case ($\gamma<0$)
is obtained simply by inverting the energies ($E_{\nu}{\rightarrow}-E_{\nu}$).
The eigenfunction characteristics described in Figures \ref{3bosonedgestates}
and \ref{fig_size-dependence} are identical in the repulsive case.


\begin{figure}
\includegraphics[width=0.95\columnwidth]{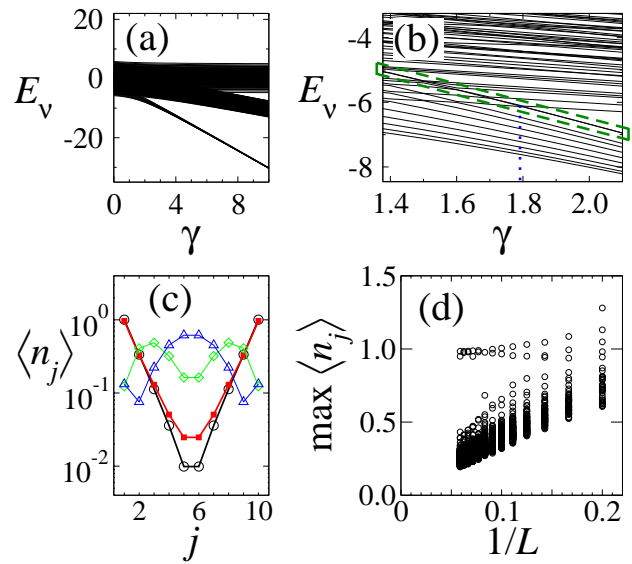}
\caption{ \label{fig_gammadependence}
(Color online.)  Spectrum and density profiles at smaller interactions, for
  three-boson system in 10 sites ($t=1$).
(a) The bands merge at small enough $\gamma$.
(b) For this system the breather band loses its identity around
$\gamma\sim1.8$.  We follow the edge-localized states at smaller $\gamma$;
shown surrounded by a thick-lined green polygon.  At smaller $\gamma$, the
$\nu=(L-1)$ and $\nu=L$ states are no longer the edge states.
(c) Site occupancies at  $\gamma=1.4$.  The almost exponentially localized
curves are for the edge states, and the two non-localized curves are the
$\nu=(L-1)$ and $\nu=L$ states, which are no longer the edge-localized states.
(d) Size-dependence of largest site occupancy. 
}
\end{figure} 


\emph{Weak interactions}. ---
%
%
It is natural to ask whether the edge-localization survives for weaker
interactions.  In particular, for three bosons and $\gamma\lesssim{2}$, the breather band
merges with the $2+1$ band (Figure \ref{fig_gammadependence}-a).
One might speculate that there might be a sharply defined value of $\gamma$
below which there is no localization at the edges.

We address this question by following the two edge states adiabatically to
lower values of $\gamma$, for three bosons in 10 sites, as shown in Figure
\ref{fig_gammadependence}-b. Note that, after the breather band has merged,
the $\nu=L-1$ and $\nu=L$ states are no longer the edge states.  In Figure
\ref{fig_gammadependence}-c we plot the spatial density profiles of the two
edge states, as well as the non-localized $\nu=L-1$ and $\nu=L$ states.
Figure \ref{fig_gammadependence}-d shows the size dependence of
$\max\xpct{n_j}$.  The localization is weaker than in the large-$\gamma$
situation ($\max\xpct{n_j}$ is less than 1.5 and the exponential decay is
imperfect), but it is still present.  Thus the localization phenomenon does
not completely disappear at some sharp value of $\gamma$, at least for small
($L\sim{10}$) lattice sizes.
%

\section{Perturbation theory}

The basic explanation for the edge-localization phenomenon is that the
edge states split off from the rest of the breather band, for
three or more particles.  We will now explain this energy splitting through
perturbative calculations in the hopping parameter.
It is helpful to introduce an effective single-particle model, which contains
only the parts of the Hilbert space relevant for this analysis.

\begin{figure}
\includegraphics[width=0.95\columnwidth]{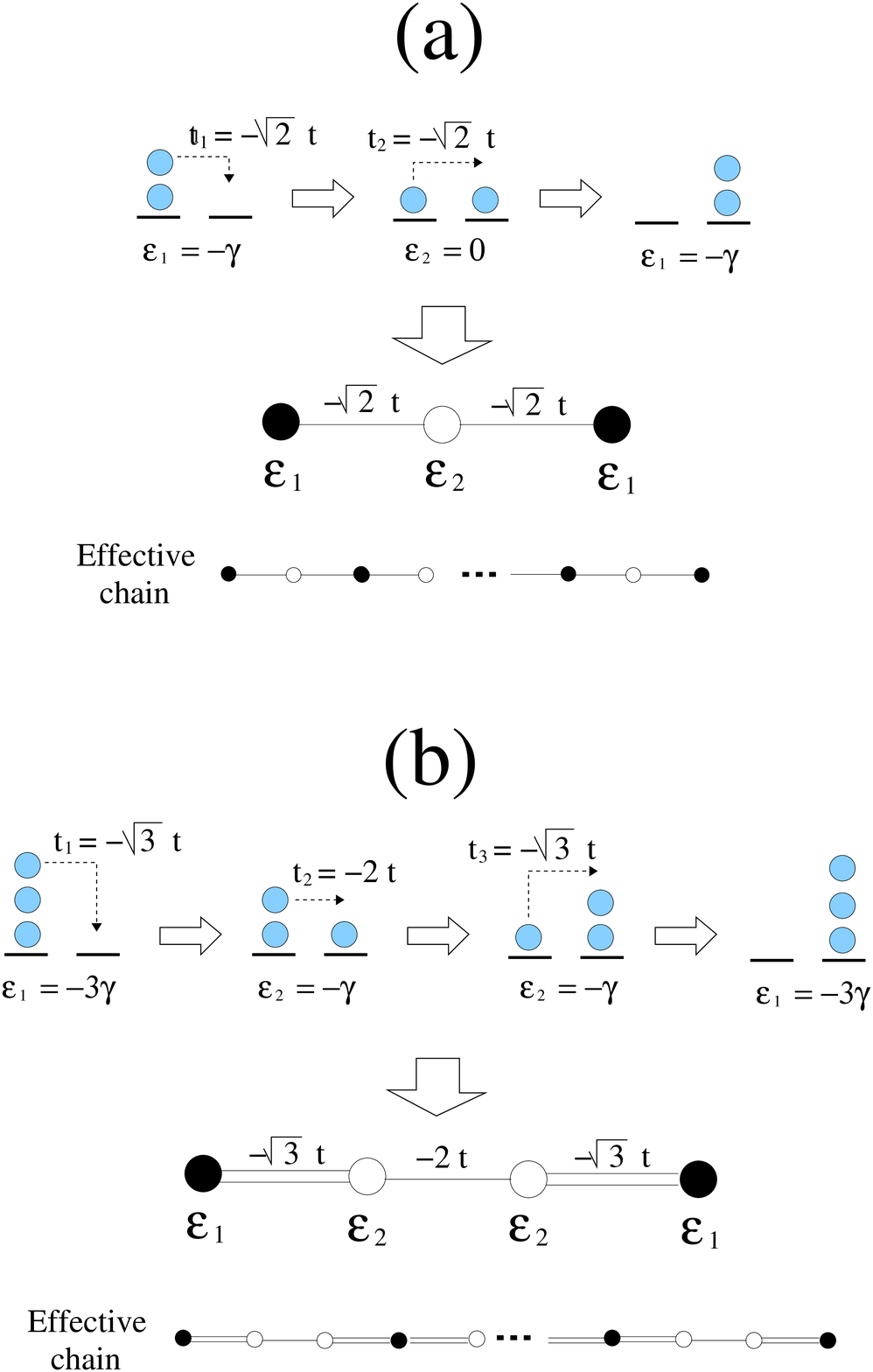}
\caption{\label{fig_effective-single-particle} 
(Color online.)  Construction of the effective single-particle chain for the
Bose-Hubbard model with (a) two bosons and (b) three bosons. The sequence of
hops of the two bosons (grey circles), initially at the same lattice site, are
shown with the corresponding values of the matrix element of the Bose-Hubbard
Hamiltonian. The energies $\varepsilon_1$ and $\varepsilon_2$ of the states
after each hop are also shown.  The resulting effective single-particle chain
is shown in the lower part of the figures.}
\end{figure}

For the two-boson case, we keep the $L$ breather band states, each
corresponding to both bosons in the same site.  We keep only those states of
the continuum that are necessary as intermediate configurations to go from one
breather state to another, \emph{i.e.}, states where the two bosons are in
neighboring sites.  In Figure \ref{fig_effective-single-particle}-a, the
2-boson states are shown with filled circles and the intermediate 1+1-boson
states as open circles; together, they form the single-particle effective
tight-binding chain.  This effective chain has two different on-site energies
$\varepsilon_1=-\gamma$ and $\varepsilon_2=0$ alternating along the chain
(Figure \ref{fig_effective-single-particle}-a).  The effective hopping strength
is $\sqrt{2}t$, since the hopping is always from or to a doubly occuppied site
in the original model.  For a fixed size $L$ of the Bose-Hubbard chain, the
size of the effective single-particle chain, \emph{i.e.}, the number of states
retained from the original Hilbert space, is $L+(L-1) = (2L-1)$.

Similarly, for the three-boson case we retain the $L$ breather band states,
and
only those states of the (2+1) band that are necessary as
intermediate configurations to go from one breather state to another,
\emph{i.e.}, states where the two bosons and the lone boson are in
neighboring sites.
The single-particle chain now has two on-site energies
$\varepsilon_1=-3\gamma$ and $\varepsilon_2=-\gamma$, and two hoppings
$t_1=\sqrt{3}t$ and $t_2=2t$. They form a chain of basis three per unit cell
(Figure \ref{fig_effective-single-particle}-b).


\emph{Degeneracy splitting} ---
At zero hopping ($t=0$) or infinite interaction, all bands, including the
breather band which we are particularly interested in, are perfectly
degenerate.
We analyze the splitting of the breather band spectrum perturbatively in the
hopping parameter $t$.  The Hamiltonian is $\hat{H}= \hat{H}_0 + t\hat{H}_1$,
where $\hat{H}_0$ is the interaction term.  
Since the relevant states of the Hilbert space are arranged along a single
chain in the single-particle effective picture, we can use the position on the
effective chain as a label for the relevant states.
For example, for the two-boson system,
\begin{equation}
\hat{H}_0 = \sum_{m=1}^{L} \varepsilon_1|2m-1\rangle\langle 2m-1| +
\sum_{m=1}^{L-1} \varepsilon_2|2m\rangle\langle 2m|  \; ,
\end{equation}
while the perturbation $\hat{H}_1$ contains hopping terms like
$\ket{2m}\bra{2m+1}$.


For two bosons, the lowest order at which the perturbation has nontrivial
effects is $\mathcal{O}(t^2)$.  The hopping perturbation at this order already
connects the states of the ground state manifold to each other, leading to a
complete lifting of the degeneracy, so that the non-degenerate breather band
emerges.  The split energies are approximately
$E_{2m-1} \approx -\gamma -4t^2(1+{\cos}k_{m})/\gamma.$

For the three-boson case, one can carry out a similar analysis.  The
crucial difference is that, at second order in $t$, the
lowest-manifold states are each connected to themselves, and thus
receive an energy shift, but are not connected to other states within
the manifold, and therefore the degeneracy is not  lifted.
The energy shifts are different for the edge and non-edge states,
because there is a single path for an edge state to couple to itself
via two hopping events, while each non-edge state has two such paths.
This is visually obvious through hopping events in real space in our
effective single-particle chain (Figure
\ref{fig_effective-single-particle}).  The shifts at second order are
\begin{equation} \label{eq:perttheor}
\begin{split}
E_{\rm edge} = -3\gamma - \frac{3}{2\gamma} t^2 \, , \\
E_{\rm non-edge} = -3\gamma - \frac{3}{\gamma} t^2 \, .
\end{split}
\end{equation}
%
At the next order ($t^3$) the degeneracy of the non-edge states is lifted,
since three hopping events are required to connect two distinct members of the
breather manifold.
The degeneracy of the two edge states is only broken at much higher order.

This analysis reveals the reason for the separating out of the edge states
from the rest of the breather band.  The energy shift of the edge states
happens at lower order in the hopping than the order at which the degeneracy
of the breather band is lifted. Therefore the edge states are robustly
separated out for large interactions.

\begin{figure}
\includegraphics[width=0.95\columnwidth]{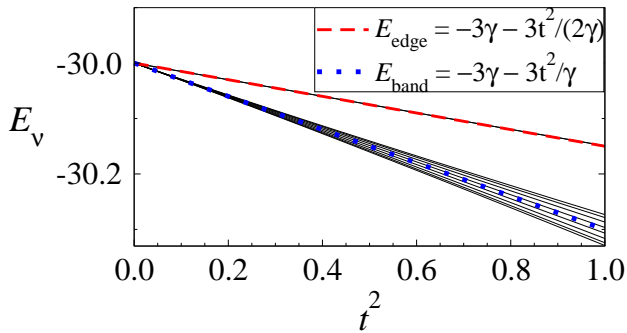}
\caption{\label{spectrumN31hopping}
(Color online.)  Energies from second-order perturbation theory (dashed and
  dotted lines) compared with full spectrum (solid lines), for $\gamma=10$ and
  $L=10$.
%
%
}
\end{figure} 

Figure \ref{spectrumN31hopping} plots the second-order perturbative results
for the breather band energies, comparing with the exact energies computed
numerically.  The splitting of the edge-localized states from the rest of the
band, which is the essential issue here, is well described by perturbation
theory.  The degeneracy lifting of the non-edge breather states is not
captured in the second-order expressions.




\section{Four or more bosons} \label{sec_4bosons}

\begin{figure}[!t]
\includegraphics[width=3.3in]{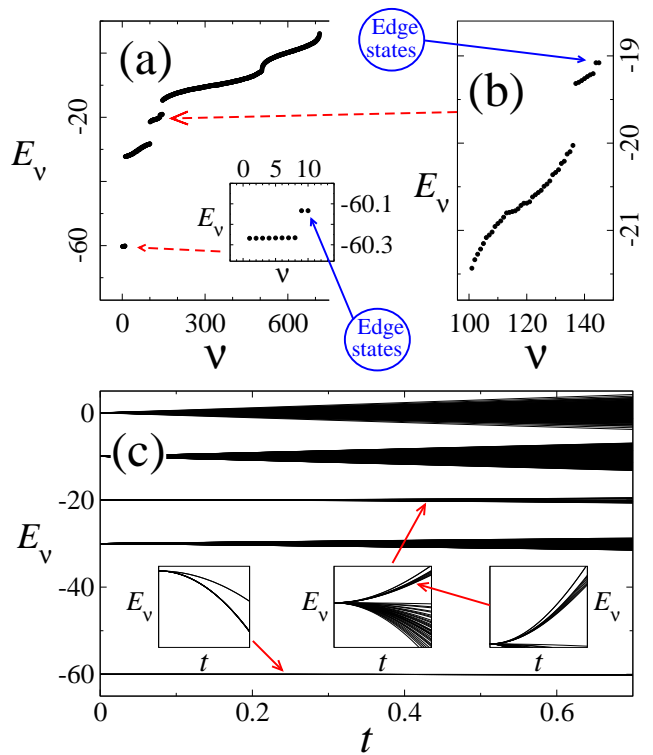}
\caption{\label{fig_4bspectra} 
(Color online.)  (a) Energy spectrum of the Bose-Hubbard chain with four
  bosons, plotted against eigenvalue label.  The inset zooms onto the breather
  band, while (b) zooms onto the (2+2)-boson band, as indicated by the dotted
  red arrows.  Here $\gamma=10$, $t=1$, and $L=10$. (c) Spectrum of same chain
  as a function of the hopping parameter $t$.  The insets focus, from left to
  right, on the 4-boson bound state band, the 2+2-boson band, and the
  2+2-boson bound state sub-band, showing the separation of edge-localized
  states out of the former and the latter.
}
\end{figure} 

\begin{figure}[!t]
\includegraphics[width=0.95\columnwidth]{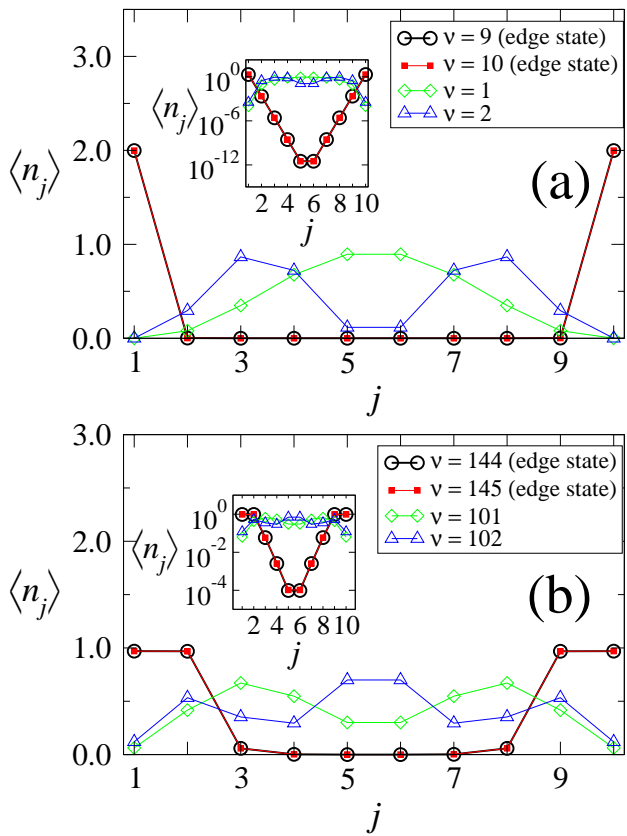}
\caption{\label{fig_4bedgestates} 
Profile of the site occupancy in the Bose-Hubbard chain with four bosons for
four eigenstates in (a) the 4-boson bound state band and (b) the 2+2-boson
bound state band. The same plots in semilog scale are shown in the insets,
showing exponential localization. Here $L=10$ and $\gamma=10$.
}
\end{figure} 

Edge-localized eigenstates also exist in Bose-Hubbard chains with $n{\geq}3$
bosons; the perturbative argument we provided for 3 bosons can readily be
extended to the general case.  At strong interactions, the lowest band is the
breather band with $n$-boson bound states.  At $t=0$, this band is collapsed
as the $L$-fold degenerate ground state.  For $t\neq0$ the degeneracy is
lifted at order $\mathcal{O}(t^n)$, but the edge states already have a
distinct energy shift at $\mathcal{O}(t^2)$, leading to edge localization.  As
in the $n=3$ case, one can also visualize the perturbative calculation with an
effective single-particle model retaining only the relevant basis states.
This will now have $n$-site unit cells, \emph{i.e.}, sites representing
breather states separated by $n-1$ sites representing states from other bands.
 

In addition to the edge-localized states with all $n$ bosons situated at the
edge, for $n>3$ the open Bose-Hubbard chain also has edge states with more
complicated structure.  We demonstrate this for $n=4$ in Figure
\ref{fig_4bspectra}.  Other than the two edge modes on top of the breather
band, we see features in the 2+2 band.  As in the translation-symmetric case
\cite{Dorignac2004PRL93}, this band has a sub-band separating out (middle
inset in Figure \ref{fig_4bspectra}-c).  This subband of $L-1$ states is
characterized by two doubly-occuppied sites neighboring each other (2+2 bound
states), while the rest of the 2+2 band is dominated by two occuppied sites at
larger distances from each other.  Unlike the translation-symmetric case,
however, if one zooms in further onto this sub-band, two edge-localized states
separate out (rightmost inset in Figure \ref{fig_4bspectra}-c).  These edge
states have the structure of two bosons at the edge, and the other two at the
next-to-edge site.  Figure \ref{fig_4bedgestates}-a and
\ref{fig_4bedgestates}-b show the density profiles of these two types of edge
states in the $n=4$ case.

The above results demonstrate the existence of more and more complicated
additional edge states as the number of particles is increased.

\section{Spinless fermions}

\begin{figure}
\includegraphics[width=0.95\columnwidth]{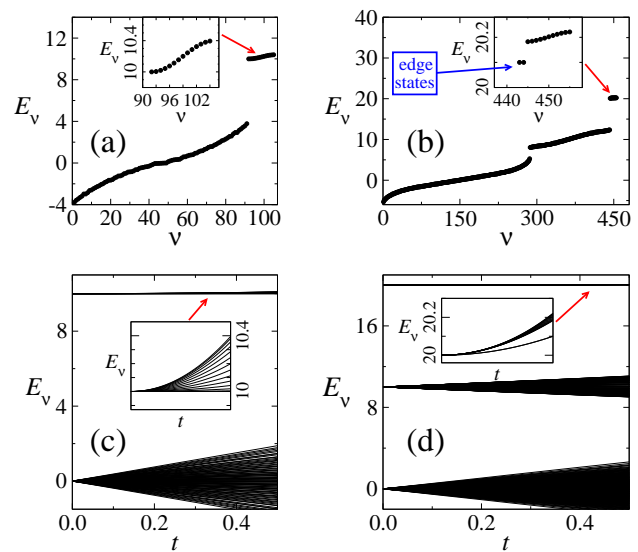}
\caption{\label{fig_spectrumSpinlessFerm}
(Color online.)  Energy spectrum of the spinless fermion model, for two
fermions in the left panels (a,c) and three fermions in the right panels
(b,d).  Here $L=15$ and $V=10$. Insets focus on the (topmost) breather band.
The three-fermion cases (b,d) have two states separating from this band; the
two-fermion cases (a,c) do not have this feature. 
}
\end{figure} 

We now turn to the spinless fermion model described by the Hamiltonian
\eqref{eq:tVmodelHamiltonian}. 
Figure \ref{fig_spectrumSpinlessFerm} displays through numerically calculated
spectral properties that in this model, edge states do not exist for the
two-fermion case, but appear when there are three or more fermions.  The
situation is thus similar to the Bose-Hubbard model.  

We first note that the spectrum of this model contains a breather band as in
the Bose-Hubbard model \cite{EilbeckPhysicaD78}.  Fermi statistics forbids
multiple occupancy of the sites; so the breather modes correspond to all the
fermions clustered in a connected segment of the lattice.  Since we are using
repulsive interactions ($V>0$), the breather band now appears at the top of
the spectrum.

The spectral splitting of edge states is analogous to what we have described
in the Bose-Hubbard case.  For the two-fermion case, the breather band is
completely split because the separation for edge and non-edge states all
occurs at the same (second) order in $t/V$.  For three or more fermions,
however, the two edge states split off from the main band at lower order than
the degeneracy-lifting of the rest of the band.  As a result, there are now
two edge-localized states separated at the bottom of the breather band.

\begin{figure}
\includegraphics[width=0.95\columnwidth]{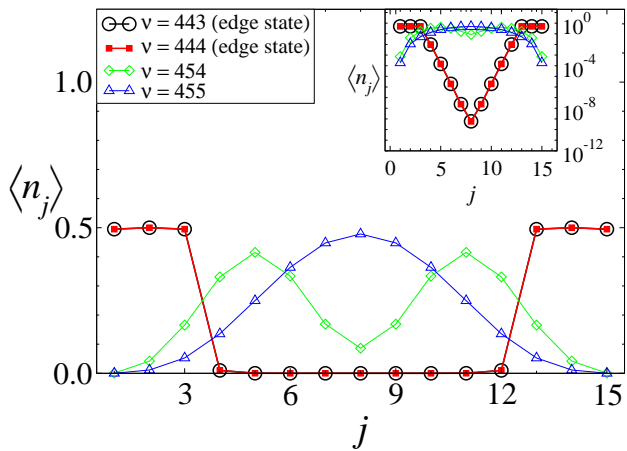}
\caption{\label{3fermionedgestates}
(Color online.)  Spatial profile of fermion occupancies, for several
eigenstates of the three-fermion model with $L=15$ and $V=10$.  Same plot in
semilog scale is shown in the inset, indicating exponential localization.
}
\end{figure} 

Figure \ref{3fermionedgestates} shows density profiles for the edge-localized
states as well as some non-localized states from the breather band.  Each edge
state now has a `width' of three sites, because of fermionic statistics
forbidding double occupancies.  
Since the two eigenstates are predominantly linear combinations of left-edge
$\ket{1110000...}$ and right-edge $\ket{...0000111}$ states, they have
occupancies of $n_j{\approx}1/2$ at the last three sites of each edge.  As in the
Bose-Hubbard case, the logarithmic plot makes clear the exponential nature of
the localization.  In this case the exponential decay starts after the third
site.

\begin{figure}
\includegraphics[width=0.95\columnwidth]{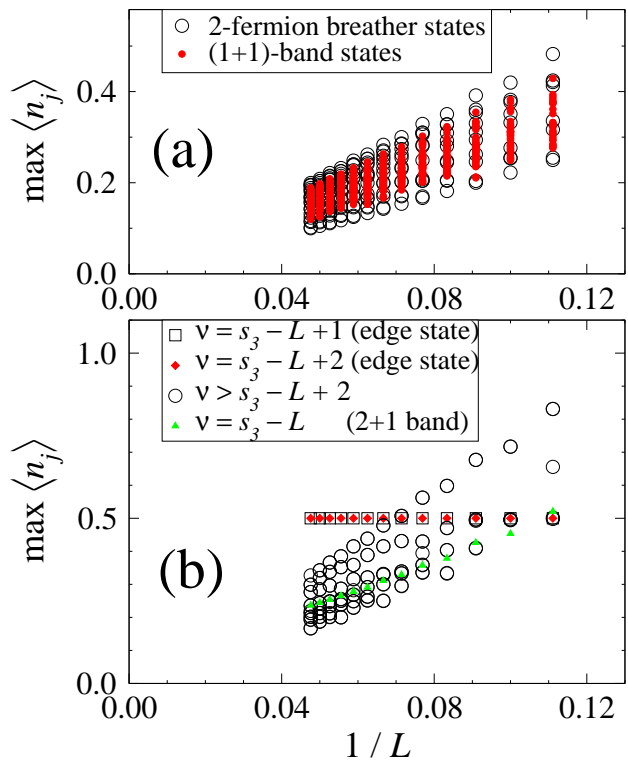}
\caption{\label{fig_spinlessFerm_L-dependence} 
(Color online.)  $1/L$-dependence of $\max\xpct{n_j}$ for (a) the eigenstates
of the two-fermion chain, and (b) different eigenstates of the three-fermion
chain. Here $V=10$, and $L$ goes from $L=9$ to $L=21$. Here $s_3$ is the
Hilbert space size.
}
\end{figure} 

In Figure \ref{fig_spinlessFerm_L-dependence}, we display the robustness of
the 3-fermion edge states by plotting the size-dependence of the maximum site
occupancy, for a fixed interaction potential.  Analogous to the Bose-Hubbard
model (Figure \ref{fig_size-dependence}), the two-fermion case has
$\max\{\xpct{n_j}_{\nu}\}$ values all varying as ${\sim}L^{-1}$, for all
states.  In the three-fermion case, all eigenstates except two also have
linear $1/L$-dependence, while the two exceptions are the edge states for
which $\max\{\xpct{n_j}_{\nu}\}$ has the constant value of 1/2.  This
demonstrates, once again, that all states in the two-fermion case are extended
in space, while the three-fermion system possesses two localized states.

\begin{figure*}
\includegraphics[width=0.7\textwidth]{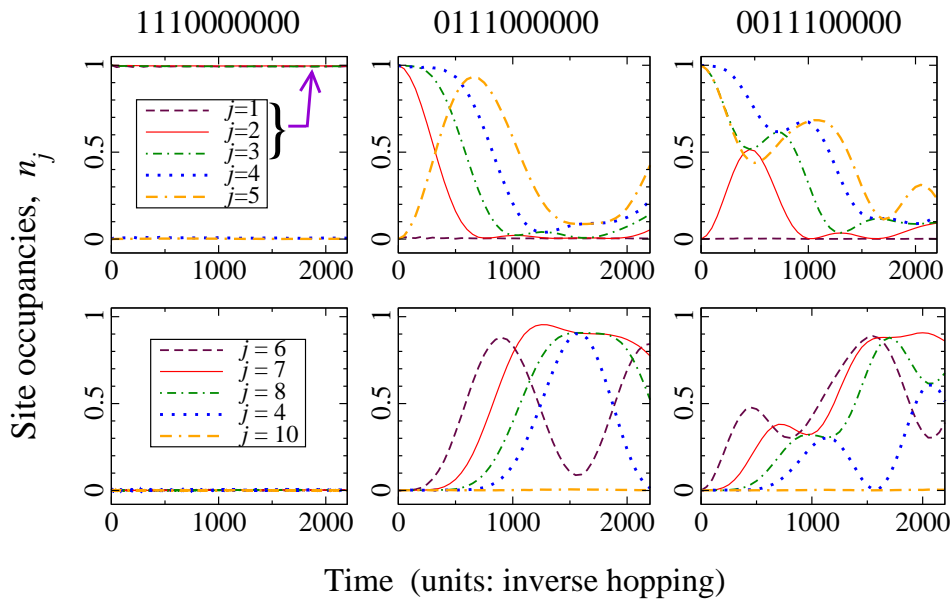}
\caption{\label{fig_3fermions10sites_dynamics} 
(Color online.)  Dynamics of three fermions in ten sites, shown through the
evolution of site occupancies $n_j$.  The three different initial conditions
are shown on top.  In each case, to avoid clutter we split the ten $n_j$ into
two groups of five.  Thus the left top and left bottom describe two parts of
the same evolving chain.  Here $V=20$ and $t=1$.
}
\end{figure*} 

\section{Dynamics (time dependence)}  \label{sec_dynamics}

It is instructive to study the localization phenomenon through explicit time
evolution calculations.
For the Bose-Hubbard model, some dynamical results have appeared in Ref.\
\cite{Pouthier_PRB2007},
In Figure \ref{fig_3fermions10sites_dynamics}, we present temporal dynamics
results for the spinless-fermion model.

For a three-fermion system in $L=10$ sites, we present the evolution of
the occupancies of individual sites, $n_j(t)$, after the system is started with
the three fermions in consecutive sites.
The left panels display the behavior when the fermions in the initial state
are concentrated at one edge (the first three sites).  There is no appreciable
dynamics at the time scales shown, in marked contrast to the cases where the
fermions start off at three other consecutive sites (middle panels and right
panels).  This difference is a dynamical demonstration of the edge
localization phenomenon.

Additional physics can be gleaned from the temporal dynamics shown in Figure
\ref{fig_3fermions10sites_dynamics}.
In the cases where we start from three non-edge consecutive sites (center and
right panels), the first and last sites are never excited ($n_{j=1}$ and
$n_{j=10}$ remain practically zero).  This reflects the spectral separation of
the edge states, $\ket{e_{\pm}} \approx (\ket{1110000...} \pm \ket{...0000111})/\sqrt{2}$, from the rest of the breather band.  Since the initial
configuration is within the breather band, the dynamics is dominated by this
band. 
Since the edge-localized subspace spanned by the basis $\{\ket{1110000...},\ket{...0000111}\}$ is separated from the rest of the breather band, the last/first sites are not excited.

\emph{Time scales}. ---
Figure \ref{fig_3fermions10sites_dynamics} highlights dynamics at the scale of
hundreds to thousands of $t^{-1}$ units.  The reason is that the dominant
dynamics for our chosen initial conditions is that within the breather band;
hence we are interested in coherent hopping of the three fermions.  Such a
cooperative hop event of three fermions, from one soliton (neighboring)
configuration to the next, involves two intermediate states that are
energetically $\sim{V}$ away, in the (2+1) band.  Thus the energy involved is
of order $t^3/V^2$, so that the time scales are ${\sim}V^2/t^3$.  For $V=20$,
this leads to the above-mentioned time-scales.  Performing simulations at
other values of $V$, we have seen that the relevant time scale indeed varies
as ${\sim}V^2$.

There is, of course, additional dynamics at other time scales.  Tiny
high-frequency wiggles can be seen in our data at scales of ${\sim}t^{-1}$,
representing high-energy inter-band processes.  Also, considering the left
panel of Figure \ref{fig_3fermions10sites_dynamics}, we note that the state
$\ket{1110000...}$ is not itself an eigenstate; the true eigenstates are
$\ket{e_{\pm}}$.  Thus, there will be oscillations involving these two edge
states, which is not visible here because the relevant time scales are much
higher, and grow exponentially with the chain length.

\section{Conclusions} \label{conclusions}

In this article, we have presented a straightforward and natural mechanism for
localization in quantum lattice systems, namely the existence of open boundaries.  Our
edge-localization is a cooperative, as opposed to single-particle,
phenomenon.  
We have provided perturbative arguments to explain the energy spectrum
structure that lies at the heart of this localization phenomenon, thereby also
explaining why at least three particles are required for the localization.  
In addition, we have showed the appearance of richer edge configurations that
appear for larger numbers of particles, and characterized the energy spectrum
and localization through a study of explicit temporal dynamics.

\emph{Energy scales}. --- At strong interactions, the bands are the most
pronounced feature of the energy spectrum.  The band energies are set by the
interaction strength ($U$ or $V$).  One can then think of finer features of
these bands, in various orders of the hopping $t$.  There is always an energy
shift at second order in $t$, since a basis state can be connected to itself
through two hops.  Whenever the degeneracy lifting requires higher than second
order in $t$, we can have sub-band structures due to differing energy shifts
at second order.
In the two-boson or two-fermion case, this possibility does not exist.  In the
three-boson or three-fermion case, only the breather band splits at higher
than second order, while the other two bands have degeneracy lifting at linear
order.  (The linear splitting is manifested in the shapes of the non-breather
bands in Figures \ref{fig_BH-spectra}-d and \ref{fig_spectrumSpinlessFerm}-d.)
Therefore in the three-particle systems only the breather band can have a
subband; this subband turns out to have two members which are edge states. 

To rephrase, edge localization from the breather band can be seen as a result
of the competition between (a) the distinct energy shifts $\sim{t^2}/U$
recieved by the edge states, and (b) the degeneracy splitting for which the
energy scale is $\sim{t^n}/U^{n-1}$, for an $n$-boson system.  For $n>3$,
analogous energy scale competitions can also play a role in other
(multi-breather) bands where splitting occurs at larger order, leading to
subbands and more complicated edge states.  We have illustrated this for
$n=4$, in Section \ref{sec_4bosons}.

Finally, we note that the two edge states separating out from the breather
band are themselves very nearly degenerate at strong interactions.  This
degeneracy gets broken only at much higher order; the relevant energy scale is
$U(t/U)^{\beta}$, with $\beta\sim{L}$ for a chain with $L$ sites.

Interestingly, some of these energy scale issues have been probed dynamically
in Section \ref{sec_dynamics}, through the study of temporal evolution.

\emph{Possible applications}. --- The edge localization has real-space effects
on the dynamics, as revealed by our time evolution calculations (Section
\ref{sec_dynamics} and Figure \ref{fig_3fermions10sites_dynamics}).  This
raises the possibility of exploiting these effects for experimental quantum
control of bosons or fermions in one-dimensional lattices.
If one starts at a breather-band configuration that is not an edge state, the
edge modes will not be excited, and conversely, fermions or bosons populated
in an edge configuration remain stable in that configuration for a long time
(Figure \ref{fig_3fermions10sites_dynamics}).  One can conceivably use this
effect, arising from fine structures in the energy spectrum, to manipulate and
select sites for a few-particle lattice system in cold-atom or quantum wire
experiments.

\emph{Open issues}. ---
Our work opens up a host of open questions and issues, of which we mention a
few.  First, our calculations shown in Figure \ref{fig_gammadependence}
indicate that for moderate-sized lattices, edge localization persists in some
form at weak interactions.  The fate of this weak-interaction localization for
large lattice sizes remains an un-resolved question.

Our work with two separate Hamiltonians results suggest that edge localization
is a generic phenomena for quantum lattice models.  Analogous phenomena can
possibly be found in other one-dimensional itinerant fermionic and bosonic
models.  It is obviously of interest to investigate this phenomenon in various
other quantum chain models, such as the one-dimensional fermionic Hubbard and
extended Hubbard models.

In the discrete nonlinear Schr\"odinger equation, one can find localized
states at all lattice sites, both bulk and edge
\cite{MolinaVicencioKivshar_OptLett2006}.  In contrast, for three or four
bosons we have only found localization of all particles at the edge, even
though some of our localized states have a finite width of more than one site.
Can the Bose-Hubbard model support solitons some distance away from the edge,
perhaps at larger boson numbers?

Finally, since edge-localized classical breathers are known for
two-dimensional lattices \cite{VicencioFlachMolinaKivshar_PhysLett07}, there
is the intriguing possibility of edge states in two-dimensional quantum
lattices, which remains unexplored at present.

\acknowledgments

The authors would like to thank A.~L\"auchli and Jean Pierre Nguenang for
discussions. This work was supported by the ESF network-programme AQDJJ.

\end{document}